# On the Composition of the Long Tail of Business Processes: Implications from a Process Mining Study

Marcus Fischer, Adrian Hofmann, Florian Imgrund, Christian Janiesch, Axel Winkelmann

**Abstract:**

*Digital transformation forces companies to rethink their processes to meet current customer needs. Business Process Management (BPM) can provide the means to structure and tackle this change. However, most approaches to BPM face restrictions on the number of processes they can optimize at a time due to complexity and resource restrictions. Investigating this shortcoming, the concept of the long tail of business processes suggests a hybrid approach that entails managing important processes centrally, while incrementally improving the majority of processes at their place of execution. This study scrutinizes this observation as well as corresponding implications. First, we define a system of indicators to automatically prioritize processes based on execution data. Second, we use process mining to analyze processes from multiple companies to investigate the distribution of process value in terms of their process variants. Third, we examine the characteristics of the process variants contained in the short head and the long tail to derive and justify recommendations for their management. Our results suggest that the assumption of a long-tailed distribution holds across companies and indicators and also applies to the overall improvement potential of processes and their variants. Across all cases, process variants in the long tail were characterized by fewer customer contacts, lower execution frequencies, and a larger number of involved stakeholders, making them suitable candidates for distributed improvement.*

# On the Composition of the Long Tail of Business Processes: Implications from a Process Mining Study

## 1     Introduction

Digital transformation embeds companies in a fast-changing and digitized economy that leverages information technology (IT) [1]. It constantly forces companies to maintain a comprehensive understanding of their processes and operations to adapt and evolve their organizations [2]. A process-oriented structure supports them in identifying customer touchpoints, determining workflows, and specifying data requirements in different parts of the value creation process [3, 4]. Generally perceived as a means for achieving streamlined operations, cost reductions, and improvements in quality and productivity, the concept of business process management (BPM) has gained importance in recent years [5-7]. Research and practice have introduced various approaches with different purposes and outcomes [8].

At their core, these concepts assume that a process's value and, thus, its need and potential for constant improvement depends on different characteristics, including its importance, health, and feasibility [7, 9]. Factors such as increasing project complexity and skill requirements limit the economic feasibility of improvement projects to managing only a handful of processes at a time [7]. Processes are prioritized and selected accordingly. The unrealized improvement potential has not been quantified.

Based on Anderson's theory of long tail economics [10], Imgrund, et al. [11] hypothesize that the improvement potential of processes follows a long-tailed distribution if determined for all processes of an organization [cf. also 12]. This implies that companies possess a few processes with significant potential or need for improvement that together form the short head of the process distribution. Beyond that, a long tail contains lower-value processes that are typically not identified as suitable optimization candidates and thus neglected [11]. The theory entails that the sum of improvement potential residing in the long tail is not neglectable as understood by the Pareto principle but represent significant unrealized value.

Despite its far-reaching implications for the practice of BPM, Imgrund et al.'s assumption of the existence of a long tail has only been verified qualitatively in a single qualitative case study [13]. It has not yet been empirically analyzed across multiple companies using automated means of process mining. Furthermore, it has not been investigated if the long tail assumptions also hold true for management of variants of processes.

By defining and indicators to measure process importance, health, and feasibility automatically from real-life process event logs of three companies, we aim to test if the assumption of long-tailed distributions holds across different cases and environments using a automatable, quantitative analysis of process variants. Furthermore, we analyze the characteristics of items in the short head and the long tail of the companies' process distributions. We discuss the results in light of Imgrund, et al. [14]'s recommendations to manage the long tail of business processes. We summarize our research questions (RQ) as follows:

> *(RQ1) How can we automatically quantify the importance, health, and feasibility of business processes or their variants from process execution event logs?*
>
> *(RQ2) How do these indicators distribute across the processes or process variants of a company and do they show a long tail of processes?*
>
> *(RQ3) What are the characteristics of processes or process variants residing in the short head and the long tail of processes and what does this imply for their management?*

We answer these RQ on the microscopic level of process variants by employing methods for automated process discovery and analysis from the field of process mining [15]. The paper is structured as follows: Section 2 introduces conceptual foundations and summarizes related work. We present our research method in Section 3 and define indicators for measuring process importance, health, and feasibility in Section 4. Section 5 summarizes the results of our analysis when applying these indicators to the companies' process execution data. In Section 6, we discuss our findings, derive constitutional characteristics, and draw conclusions for their management in BPM. We conclude with a summary of findings, limitations, and future research potential.

## 2  Fundamental Knowledge and Related Work

In the following, we introduce the theory of the long tail of processes as our conceptual foundation. We continue to highlight relevant research on process mining, specifically process discovery, process performance management, process prioritization, and related work from the Business Process Intelligence (BPI) challenges that center on our datasets.

### 2.1  Theory of the Long Tail of Business Processes

BPM comprises a body of methods, techniques, and tools to identify, discover, analyze, redesign, implement, and monitor business processes in according process lifecycle stages [7]. Automated processes are executed by process-aware information systems typically on the basis of process models. These systems also manage involved applications, people, and information [7].

Most companies organize BPM as a central initiative that pools resources, expertise, and competencies in some sort of BPM center of excellence to yield benefits from specialization and learning effects [5, 7]. However, as the scope of central BPM is naturally limited by resource constraints, companies must explicitly prioritize processes that yield the highest expected benefits, while neglecting those offering only little direct impact on their business success [7].

To conceptualize this phenomenon, Imgrund, et al. [11] have repurposed Anderson [10]'s long tail economics and introduced the theory of the *long tail of business processes*, which defines BPM as a profit-maximization problem. It implies that companies seek to maximize their profit $P$ as a function of BPM surpluses $E[BPM(x)]$ and costs $C(x)$. The expected surplus denotes the benefit of structured process improvements. By contrast, the complexity of a company's operations, inconsistencies in functional decision making, and the general erosion of efficiency in large projects reduce the surplus of BPM. Besides setup costs for employee training and IT support, BPM yields direct costs for managing a particular process that are linked to its degree of standardization and the existence of learning effects through knowledge spillovers.

In the most part for economic reasons, companies manage all processes that have an advantageous proportion of expected surpluses to costs. The *line of manageability* denotes the process at which both determinants break even. It is historically associated with the Pareto principle. We graphically summarize the theory of the long tail of business processes in Figure 1. van der Aalst, et al. [12] also refer to this concept to illustrate the applicability of robotic process automation for the middle-part of the long tail of processes. The RQs of our research focus on prioritizing, that is ordering, the processes visualized on the *x*-axis by their expected surplus.

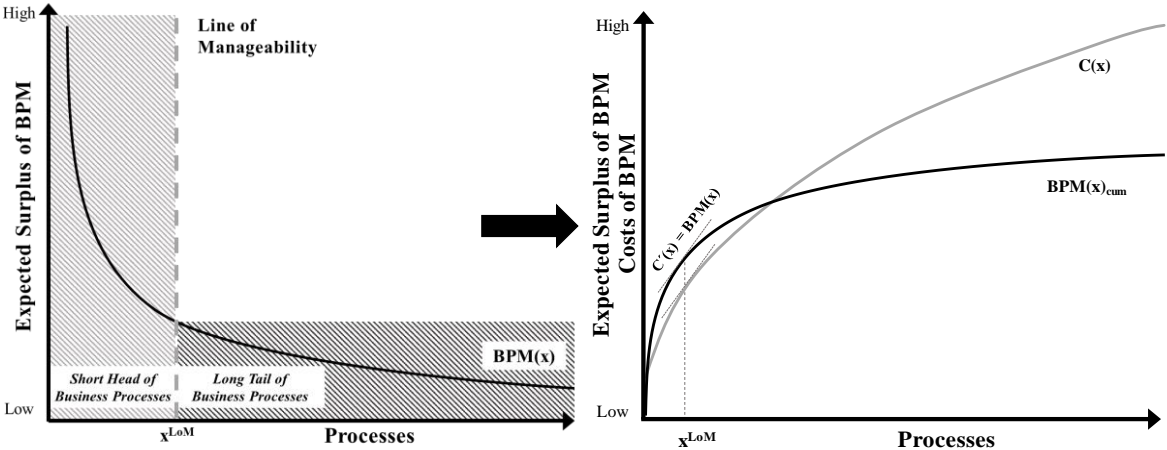

Figure 1: The Long Tail of Business Processes [11]

Imgrund et al. argue that to incorporate additional processes beyond the short head, companies must reduce their BPM costs by augmenting their central initiatives with principles of commons-based peer production [16] such as crowdsourcing or peer-to-peer collaboration [17]. They further suggest that key processes in the short head of a company's process distribution remain centrally managed, while the long tail should be continuously improved by distributed stakeholders that coordinate tasks and activities autonomously only being supervised by central resources upon request or in case of urgent alignments to the corporate objectives [11, 14].

Central BPM would then consist of a few high-skilled process analysts that constantly act upon a company's objectives and strategies. By operationalizing established methods and guidelines, these initiatives produce high-quality outcomes for various purposes, such as standardization, quality assurance, and automation. Decentral BPM builds upon multiple distributed initiatives that together form a self-organizing social system based on communication and the use of commons-based peer production. Because decentral BPM requires the participation of stakeholders, companies must provide adequate means for communication and establish process awareness and open-mindedness. Corresponding outcomes are rather part of a continuous improvement process than leading to substantial performance increases in the short term. For more details on central and decentral BPM see Imgrund, et al. [14].

Maintaining a central and decentral initiative simultaneously requires new organizational capabilities and alters all stages of the BPM lifecycle [17]. Among other aspects, this requires rules and incentives to determine the scope and outcome quality of both initiatives as well as a comprehensive indicator system to prioritize the processes [11, 14].

## 2.2  Process Mining and Process Discovery

Process mining focuses on collecting, analyzing, and interpreting data from process event logs. Event logs comprise information about process executions extracted from process-aware information systems [15], which persist messages, transactions, or modifications [18]. Among others, event logs specify names or IDs of processed business objects, involved company stakeholders (i.e. used resources), the time at which events occurred, and possibly business data [19]. Process mining is used for the tasks of process discovery, process conformance checking, and process enhancement [20].

Research has developed various techniques to implement process discovery in practice [21]. Although some processes pursue the same objective, they can involve different activity sequences. We draw upon the concept of *trace clustering*, which is suitable for decomposing event logs into homogenous subsets, for example processes or process variants [22]. Among others, bag-of-activities, hamming distance, generic edit distance, and *n*-gram models describe the most popular approaches for trace clustering [22]. The bag-of-activity approach neglects contextual information and does not account for the order in which activities are executed. Against this drawback, *n*-gram models do not only analyze stand-alone activities but also incorporate preceding and succeeding activities as an activity's context. Hamming distance operationalizes the number of character positions in which two sequences differ and is limited to traces with the same length. This problem can be mitigated with sequence alignment approaches [23]. Edit distance penalizes sequences with different lengths heavily and is not suitable for analyzing real-life event logs [22]. Against this backdrop Bose and van der Aalst [24] introduced a generalized edit distance that takes the properties of process logs into account. However, the calculation of these scores proved to be unfeasible for the larger datasets. Additionally, many internal cluster measures do not work well with custom distance measures, since cluster centers cannot be calculated trivially. Considering the benefits and disadvantages of the different trace clustering approaches, this study employs the configuration of *n*-gram models using the Euclidean distance measure.

## 2.3 Process Performance Measurement

Measuring process performance is a pivotal task of BPM and provides companies with the means to measure what they manage [25]. In addition to collecting relevant data, it entails analyzing how well a process performs compared to a set of predefined objectives and criteria. In doing so, it creates a solid foundation for prioritization decisions and supports companies in allocating organizational resources to the most promising improvement projects.

As a multidisciplinary topic, performance measurement has been extensively studied in the past [26] and authors proposed different methods, frameworks, and concepts, most prominently the balanced scorecard [27] and EFQM [28]. For a comprehensive overview of performance measurement models see van Looy and Shafagatova [26]. Despite sharing similar assumptions and objectives, these methods generally differ regarding their scope (e.g. company-level vs. process-level) and the applied measures for process performance (e.g. performance-based vs. non-performance-based) [29].

*Non-performance-based* methods entail evaluating process performance against a set of predefined criteria [30]. Companies can either define criteria based on critical success factors or by collecting information from involved employees [31]. Stakeholders are then prompted to assess process performance individually or in group workshops based on a scale that captures negative and positive perceptions toward a process [32]. By consolidating individual views, companies can obtain an in-depth understanding of the drivers and obstacles to business success. However, non-performance-based methods are typically time-consuming, complex, and prone to assessment bias [29]. While non-performance-based methods provide important qualitative evidence for process performance measurement, they are costly and unsuitable for frequently recurring analyses due to their complexity and difficulty to derive unambiguous results. Fischer, et al. [13] have used non-performance-based methods to analyze the process distribution at a small and medium-sized enterprise.

In contrast, *performance-based methods* enable companies to analyze the as-is condition of a process automatically by operationalizing process execution data. They can compare performance indicators to benchmarks or admissible value ranges. Further, they support companies to gauge the risks of process change and to track long-term developments in an automated and accurate manner [29]. The applicability of performance-based methods depends strongly upon the availability of process-aware information systems. Performance-based methods relate closely to the field of process mining, but instead of focusing on process discovery, conformance checking, and enhancement, they analyze execution data for patterns and relationships with an extract-transform-load procedure. By fostering accuracy and reproducibility, they facilitate the assessment of performance indicators, process complexity, current design flaws, and improvement potential. This study employs performance-based methods to answer our RQs.

## 2.4 Process Prioritization

Process improvement is a complex and multi-faceted topic and relies on the selection of suitable processes. Hence, process prioritization should not only target rely on cost performance but also on other criteria [7]. In the past, several concepts for measuring process performance have been introduced by research and practice [26, 29]. As a baseline approach, Davenport [9] and Dumas, et al. [7] describe the high-level criteria of *(strategic) importance*, *health* or *dysfunctionality*, and *feasibility* to assess a process's improvement potential and to guide decision making during process prioritization.

- *Importance* describes the strategic relevance of a process based on its impact on strategic goals. Hence, processes that relate more closely to key business activities and the corporate strategy are valued higher. Assessing the importance of a process, thus, requires an understanding of one's own strategic course.

- *Health* captures a process's current condition and sheds light on its conformance with predefined quality and performance expectations. It does not only highlight those processes that are particularly prone to errors but also those with low employee and customer satisfaction. The term, borrowed from medical science, makes an implicit reference to the positive assumption that normalcy is in place rather than dysfunctionality.

- Ultimately, *feasibility* describes the degree to which a process is adaptable and influenced by cultural and political constraints. Processes improvement should focus on those processes that face the least obstacles not only in terms of employee resistance to change but also in terms of technological constraints.

Although research provided companies with several performance indicators that can be related to these dimensions, they typically require adaptation to conform to different organizational contexts [25]. At industry level, Becker and Winkelman [33] define a comprehensive set of ready-to-use indicators for retail companies. Among others, similar concepts exist for the industries of manufacturing [34] and public administration [35]. However, these concepts are industry specific as they measure business-related concepts such as minimum inventory, stock depreciation, or perfect order ratio. They lack generalizability to be relevant for the analysis of process execution data independent of industry and often require business data that is not present in common event log configurations. Hence, more general indicators based on a limited number of attributes available through process-aware information systems are necessary.

Rather than specifying individual indicators to prioritize processes, Kratsch, et al. [36] introduce a data-driven approach that operationalizes different features, such as execution times and material usage, to predict the future performance of a process and to return a scheduled priority list of improvement projects. Similarly, Lehnert, et al. [29] introduce ProcessPageRank to evaluate processes against their network-adjusted need for improvement. Ohlsson, et al. [37] use two components: a process heat map and a categorization map. While the heat map serves as a blueprint for process analysis, the categorization map determines a process's current performance gap.

## 2.5  Related Work from BPI Challenges

Over the last decade, ten BPI challenges have been issued, each with a distinct dataset. Most of these datasets, such as the ones used in this work, have become common benchmark datasets for the BPM community. While the focus has shifted somewhat from year to year based to the challenge questions, certain topics have been persistent and some, such as process prediction using artificial intelligence [38], have emerged during this decade of process mining. While the BPI challenge started with questions about descriptive statistics and control flows, it has evolved into asking more complex questions, for example on customer workload prediction (2014), social network analysis (2015), or pattern mining (2016).

Related to our research, discovering the control flow of the process model(s) in the event log has been an important topic. While early contributions comprise this as a core contribution, for many analyses nowadays it has become part of the process understanding and descriptive statistics as there is sophisticated software available to (pre-)process event log data. Related to our paper, distinguishing variants has been a topic of research based on the needs of the analysis but not as a topic of its own. Traces have been grouped or clustered for example by business data [39, 40], informally by dotted charts [41], by activities [42], and by resources and process states [43]. Further, many relevant performance indicators for descriptive statistics have been proposed including execution frequency, activity duration, and resource usage. As a byproduct of other analyses, vanden Broucke, et al. [44] and Scheithauer, et al. [45] have noticed that long tail distributions exist on an individual trace level and for process variants for some of their indicators.

While instance clustering and process performance analysis have been frequent topics within the community, defining a comprehensive and balanced set of indicators for instance or variant prioritization has not yet been explored. Further, while long tail distributions have been discovered in a few submissions, an analysis of their composition has not been performed. Their significance for the management of processes in terms of a short head and a long tail of processes or variants thereof has neither been explored.

## 3  Research Design

Subsequently, we analyze process execution data from real-life event logs. As this entails processing large volumes of data from different sources, we align this research with the method of big data analytics [46]. This conforms to the view of Müller, et al. [47], who define big data analytics as the statistical

- 7 -

modeling of large and dynamic datasets of user-generated content and digital traces. We use the guidelines of Müller, et al. [47] and organize our analysis along three phases: *(1) data collection*, *(2) data analysis*, and *(3) interpretation of results.*

## 3.1 Data Collection

The concept of Imgrund, et al. [11] implies a macroscopic view of the company, where data on all processes of a company is available. This is more likely for non-performance-based measurements as performed in [13], but performance-based measurements require the availability of event log data across the company's process landscape. In a strict sense, this would entail that all processes of a company needed to be implemented in process-aware information systems, which make their data available. In practice, performance-based measurements will therefor only be able to focus on a subset of processes or require manual data collection.

Hence, as such data was not available to the authors, in this work we focus on a microscopic view of the long tail of processes and examine whether variants of a single process conform to the long tail theory. Please note that in doing so, we do not try to find a clear distinction between a process and a process variant. For the scope of our analysis, each event log contains only data of one process.

The data for this study is provided by the *4TU Centre for Research Data*, which supports research endeavors in the field of process mining with real-life and synthetic event logs. The datasets have been used for several BPI challenges and represent well-known benchmark datasets for the BPM community. Each dataset contains process execution data recorded over a period of multiple months or years. We summarize the main characteristics of the case logs in Table 1. We have selected two similar organizations (1, 2) to observe differences despite comparable focus and industry and a third for its differentness to the other two cases. All datasets comprise (internal) customer-facing processes and not pure backend processes. While in some industries, backend processes without customer interaction may be considered more valuable, we conceded that across industries this is not the case and our research must be adapted in those cases. Further, for Log 2 there are two datasets available, which are five years apart. We have analyzed both.

Table 1: Characteristics of Process Execution Event Logs

|  | **Log 1** | **Log 2** | **Log 3** |
|---|---|---|---|
| Industry | Finance | Finance | Government |
| Focus | IT service management | Loan applications | Building permit applications |
| Source | BPI 2014 [48] | BPI 2012/2017 [49, 50] | BPI 2015 [51] |

Each dataset contains process instances for one process with multiple variants. The collected datasets contain roughly 95,000 process instances with 2,176,000 events of real-life data. Each process instance has a unique identifier and between 6 and 24 labeled features. Besides numeric features, the datasets include Boolean, categorical, and text variables that are used to flag the handled business object, involved stakeholders or resources, process errors, or lifecycle phases. As a timestamp indicates when an event occurred, we can decompose process instances into an ordered list of activities. Each dataset comes with a short description of all features. Cf. Figure 2 for an example of two clustered variants from log 1. While obviously not all activities are different, there are notable difference such as *reopen* and *update* of cases and *mail to customer* and *caused by CI* in variant 2 and differences in the handling of *reassignments* and the finalization of the process from *closed* to *end*.

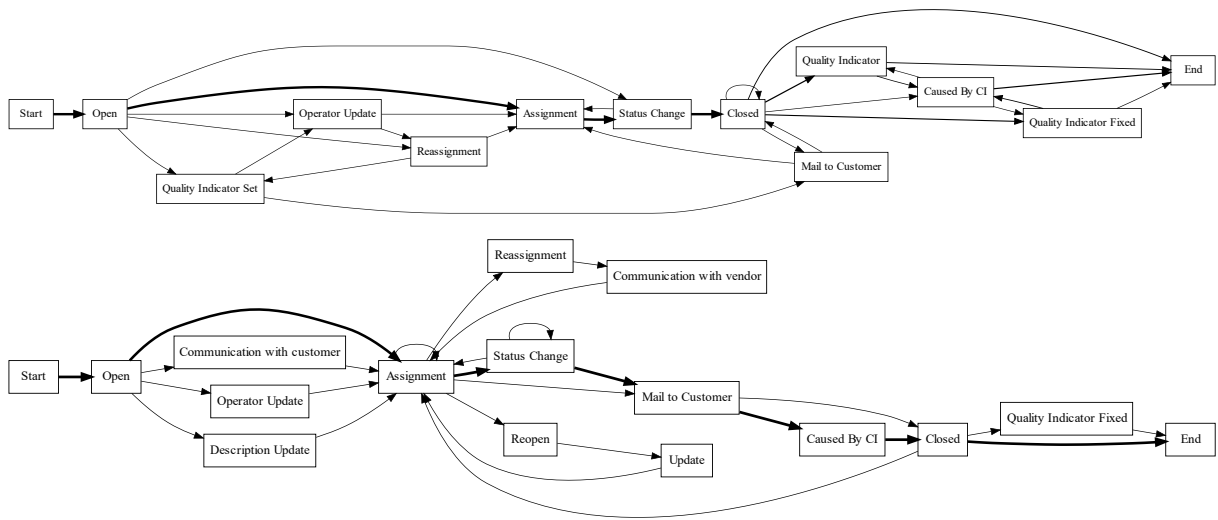

Figure 2: Comparison of Two Process Variants from Log 1: IT Service Management

## 3.2 Data Analysis

To answer the RQs of this study, we consider process performance as a broader concept that integrates a process's *importance*, *health*, and *feasibility* [9]. We operationalize these dimensions by defining a set of indicators that are supported by common event log configurations. Thereby, we ensure consistent results and facilitate the applicability of our research design beyond the borders of this study.

As the data stems from different organizations, we perform careful preprocessing to control for data quality issues, such as missing values or inconsistencies. We present multiple descriptive statistics to grasp the logs' main characteristics and employ process mining techniques for process discovery. More specifically, we use trace clustering to merge similar instances to process variants. We summarize our process mining procedure in Figure 3.

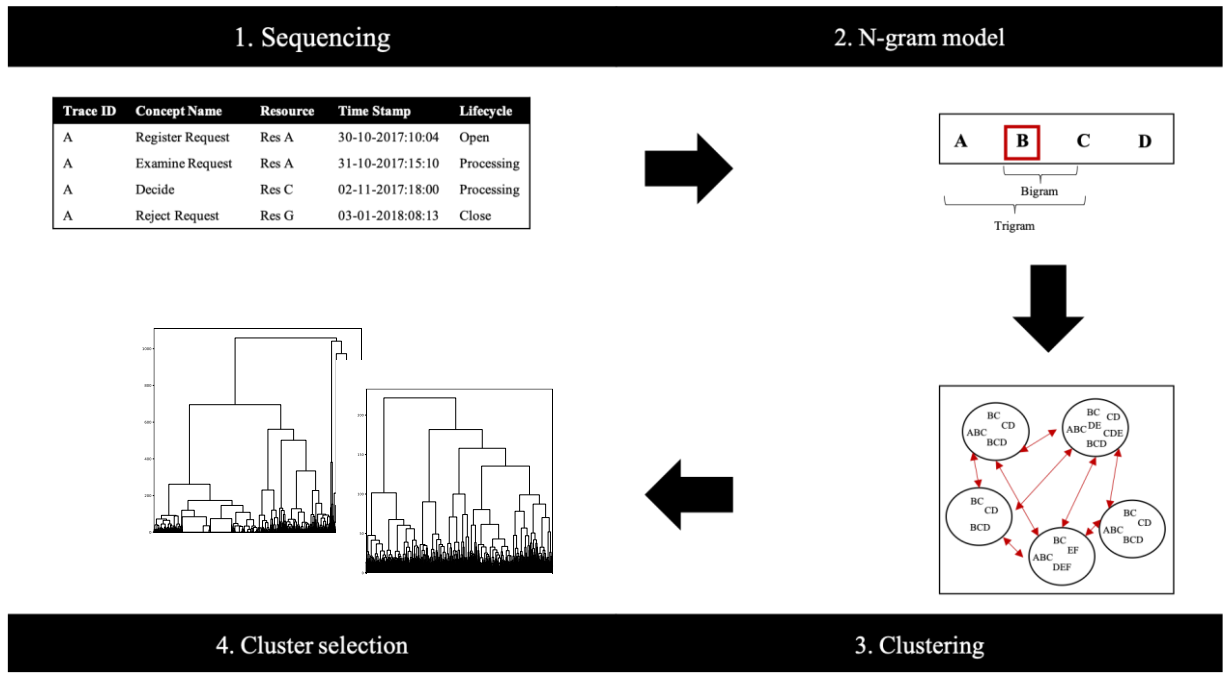

Figure 3: Process Mining Procedure Employed in this Study

We initially transform all traces to arrays of strings. Each entry corresponds to an activity in a trace. This yields a collection of *n* arrays for an event log with *n* traces. Second, we build bigram and trigram combinations to incorporate contextual information of each activity. This allows us to include the predecessors and successors of an activity and thus to account for structural properties of a process. Third, we transform the *n*-gram model into a vector space and compute the similarity of all vectors of the same event log based on Euclidian distance. We apply agglomerative hierarchical clustering based on the Ward algorithm to divide the event logs into traces. Agglomerative clustering of processes is a bottom-up procedure which starts with each process instance being its own cluster. Pairs of (similar) clusters are then merged until the final set of clusters emerges. A key advantage of agglomerative clustering is that the clustering has to be calculated only once, and from there, one can determine a suitable number of clusters by traversing the dendrogram.

Wu [52] points out that different selection criteria can yield different results and Gimpel, et al. [53] as well as Fischer, et al. [54] emphasize that research still lacks clear and proven recommendations for selecting the best measure and that human assessment is still necessary. Backhaus, et al. [55] further describe the decision between different cluster solutions as a trade-off between manageability and homogeneity. We employ a variety of internal cluster measures, including the Silhouette score, the Dunn index, and the Calinski-Harabaz score.

We use these clusters to calculate indicator distributions for each process. The indicators have been developed twofold. Based on a literature review (for selected results see Section 2), we have developed an understanding of data commonly available in process event logs and common industry-specific indicators. In total, we screened 199 papers and used 17 of them to derive indicators or indicator dimensions, most notably [56-61]. Then, we discussed these indictors with 8 process owners working in different departments and generated a shortlist of 15 process performance indicators that may have a bias towards short head processes. Hence, in addition we performed a second study by analyzing 6 robotic process automation case studies and conducting 15 interviews with vendors and consulting companies to derive 6 indicators suitable for prioritizing the long tail of processes. From these partly overlapping 21 indicators, we derived a final set of 9 indicators across the three dimensions that can be used for automated process performance analysis using event log data for both, the short head and the long tail of processes [for more details on these studies cf. 13, 62]. The implementation and data is available on B2SHARE [63].

### 3.3 Result Interpretation

Finally, we discuss our findings and analyze the resulting indicator distributions. On the one hand, we provide company-level insights by interpreting our indicators. On the other hand, we evaluate the process performance indicator curves to assess if they conform to the shape of long tail distributions as proposed by Imgrund, et al. [11]. By applying the Pareto principle to assign processes or their variants either to the short head or long tail of processes, we further examine the main characteristics of the process variants located in each of the companies' process distributions. Thereby we provide firsthand empirical insights into the composition of the long tail of business processes and justify its management in either central or decentral BPM initiatives.

## 4 Process Performance Indicators

To answer RQ1, we derive a set of indicators as a baseline approach for measuring process performance in event logs that is applicable across companies. Hence, we only rely on data that is available in common event log configurations to increase the indicators' applicability and the generalizability of the corresponding implications. That is, we refrained from using indicators that require business data attributes. We have used them to inspire the following proposal of process performance indicators. This study incorporates all three dimensions of Davenport [9] and Dumas, et al. [7] and, thereby, conforms to Kueng [64], who defines process performance as a multi-dimensional construct, which cannot be determined on the basis of a single straight-forward indicator.

We use a formal notation to define the indicators. An event log $E$ consists of a finite set of processes or process variants $P_i$. They are further composed of cases $C_{ij}$, which represent process executions. Ultimately, cases describe a selection of events $A_{ijk}$ that stand for the activities of a process. We use indices to refer to processes (*i*), cases (*j*), or activities (*k*).

## 4.1 Indicators for Process Importance

The strategic importance of a process depends on its impact on a company's business success, which is typically associated with the turnover of a process and its overall contribution to value creation. However, instead of profit increases, most BPM initiatives seek to achieve cost reductions by increasing operational efficiency and effectiveness through automation [7]. With cost-related process performance indicators, companies can obtain insights about a process's fixed and variable costs. However, execution costs (or profits) are rarely persisted in an event log. While fixed costs are independent of execution, variable costs depend on process intensity and the utilization of company resources, which can give an indication about the processes importance. Kueng [64] further argues that process importance is related to the roles and responsibilities of involved stakeholders. Consequently, in this study we view process importance from the three perspectives of *execution frequency*, *resource utilization*, and *customer interaction*.

Based on their role within a company's organizational structure, some processes are performed often, while others are executed in special cases only. Because the former type is necessary especially for handling frequently occurring events and provides essential input for other processes, it contributes to a company's value creation more significantly [65]. Therefore, we use the execution frequency ($EF$) of processes as an indicator of importance. We summarize the underlying formalism in Equation 1, whereby $|P_i|$ represents the number of process instances.

$$EF_i = |P_i|$$

Equation 1: Execution Frequency

Processes involve resources for execution. Resources such as IT systems or human workers incur cost. While we do not consider cost a sole indicator of process importance, it may be one indicator that explains some importance of the process [7]. If cost information is available, $Cost_l$ denotes the cost of resource $l$. In the absence of cost information for resources in event log data, we propose to approximate importance of resources. Resources have different roles and responsibilities. As resources in operative functions frequently perform mostly rudimentary activities, their contribution to solving a case is low. By contrast, complex customer requests require expert knowledge or special responsibilities. Involving these resources induces higher costs. Hence, corresponding processes are more important for a company's business success. Consequently, in the absence of cost information, we suggest operationalizing the importance of resources ($RU$) involved in executing processes as an indicator of process importance. We present the underlying formalism in Equation 2, whereby $Res(A_{ijk})$ denotes the resource executing activity $A_{ijk}$ and $Res^{-1}(Resource_l)$ specifies all activities executed by resource $l$. $RR(Resource_l)$ further captures the inverse relationship between the number of involvements and resource importance. Finally, process importance results as the average resource score per activity.

$$RU_i = \frac{\sum_{j=1}^{|P_i|} \sum_{k=1}^{|C_{ij}|} RR(Res(A_{ijk}))}{|C_{ij}|} \quad \text{with} \quad RR(Resource_l) = \begin{cases} Cost_l & \text{if cost avilable} \\ |Res^{-1}(Resource_l)|^{-1} & \text{else} \end{cases}$$

Equation 2: Resource Utilization

In a process landscape, important processes are typically not categorized by cost but referred to as core processes due to their relevance to customer satisfaction and impact on competitiveness and strategy success [5]. By contrast, support processes provide the capabilities of a company's day-to-day operations, without influencing customer satisfaction directly [66]. Consequently, as core processes that typically involve customer interactions are considered more important than those performed in a company's back office, we use the number of customer contacts ($CC$) as an indicator of process importance. Equation 3 summarizes the underlying formalism.

$$CC_i = \frac{\sum_{j=1}^{|P_i|} |\{A_{ijk} \in C_{ij} : A_{ijk} \text{ implies Customer Contact}\}|}{\sum_{j=1}^{|P_i|} |C_{ij}|}$$

Equation 3: Customer Contacts

### 4.2 Indicators for Process Health

Besides cost and flexibility, time and quality are typical perspectives for process performance assessments [7]. Both indicate whether or not a process conforms to predetermined specifications, for example in terms of cycle time deviations or errors. Hence, we consider them as suitable perspectives for measuring a process's health; that is the process's state of normalcy and the absence of dysfunction.

On the one hand, we employ the concept of time-related process performance used to measure a process's execution time, which describes the time necessary to handle a case. It consists of the two components: service time and wait time. While service time captures the duration of all activities required for processing a case, the concept of wait time accounts for times during which a case lies idle due to bottlenecks or other resource constraints. In most BPM approaches, process improvements relate closely to minimizing time-based indicators, such as throughput time, processing time, wait time, transportation time, or delivery time. While long execution times can be an indicator of dysfunctionality, some processes require more steps and are naturally longer. Hence, variance in execution times is a better indicator of imbalances and ailing process stability that companies focus on reducing [67]. In this study, we draw upon the two concepts of *activity duration variance* and *execution time variance* to measure time-related process health.

On the other hand, quality-related performance measures describe the degree to which a process's outcomes conform to presupposed expectations. We can distinguish between an internal and an external view with the former drawing upon employee satisfaction and perceived variance and the latter capturing customer satisfaction with either a product or the process of delivery [68]. In order to incorporate quality-related satisfaction into our indictor system, we draw upon the concepts of *execution redundancies* to measure quality-related process health.

Although processes differ regarding their role within an organizational structure, they together form an interdependent system aligned toward achieving a company's objectives and strategies. Processes do not only operationalize output from other processes but also share identical tasks and activities that are performed in different contexts. These shared activities typically involve the same requirements and workflows and, thus, should show little variation in their execution time. A high activity duration variance ($AV$) can indicate structural deficiencies due to a lack of coordination and communication. Consequently, we use the difference between the minimal and the actual execution time as a measure for process health. We present the underlying formalism in Equation 4, whereby $T_{min}(A_{ijk})$ represents the lowest execution time of an activity $A_{ijk}$ across all cases $C_{ij}$ within a process cluster $P_i$ and $T$ specifies the execution time of activity $A_{ijk}$.

$$AV_i = \frac{\sum_{j=1}^{|P_i|} \sum_{k=1}^{|C_{ij}|} \left(T(A_{ijk}) - T_{min}(A_{ijk})\right)^2}{\sum_{j=1}^{|P_i|} |C_{ij}|}$$

Equation 4: Activity Duration Variance

Processes typically include a defined set of activities and interactions. To achieve standardization and, thus, high-quality outcomes, companies must carefully analyze the requirements to establish an adequate process design for handling a range of different cases. Hence, processes or variants thereof should show only minor variance if executed repeatedly. We use the measure of execution time variance ($ETV$) to assess process health and formalize it in Equation 5, whereby $AD_i$ specifies the average duration of a process execution.

$$ETV_i = \frac{\sum_{j=1}^{|P_i|}(T(C_{ij}) - AD_i)^2}{|P_i|} \quad with \quad AD_i = \frac{\sum_{j=1}^{|P_i|} T(C_{ij})}{|P_i|}$$

Equation 5: Execution Time Variance

Processes comprise a defined control flow with several activities to handle business objects. Activities can be executed sequentially or in parallel. However, a lack of coordination and communication can yield a situation where a case is sent back and forward in multiple iterations creating unnecessary loops [69]. We draw upon these redundancies as the third indicator for process health. Equation 6 presents the underlying formalism, whereby the score of execution redundancies ($ER$) results from the number of activity tuples that occur multiple times during the execution of a process.

$$ER_i = \sum_{j=1}^{|P_i|} | \bigcup_{m=1}^{|C_{ij}|} \bigcup_{n=1}^{|C_{ij}|} \{(A_{ijm}, A_{ij(m+1)}) \in C_{ij} \times C_{ij} : (A_{ijm}, A_{ij(m+1)}) = (A_{ijn}, A_{ij(n+1)}), m \neq n\}|$$

Equation 6: Execution Redundancies

## 4.3 Indicators for Process Feasibility

Even if a process's health and its importance warrant a high process prioritization, it may not be feasible to improve a process. Dumas, et al. [7] note that a process's feasibility is chiefly influenced by its political and cultural environment. Consequently, management involvement and a general resistance towards change can hamper a process's improvement. Further, they divide feasibility-related performance into run-time and build-time flexibility. While run-time flexibility captures a process's capabilities for handling changes during process execution, build-time flexibility relates to adaptations to its design. As political and cultural perceptions are hardly persisted in an event log, we consider a process's complexity in terms of its build-time flexibility as a suitable proxy for its adaptability.

Lehnert, et al. [29] emphasize the relevance of structural and stochastic dependencies in process networks. They assess the centrality of processes within a company's organizational structure and determine their complexity and, thus, feasibility based on the number of links to other processes. To assess a process's feasibility, we draw upon the three concepts of *shared activity contexts*, *stakeholder involvement*, and *process length* to measure the complexity of a process. In contrast to the above metrics, processes are typically more feasible to manage if they are less complex. Hence, all metrics in this dimension use inverse relations or minimums.

Processes build upon sequences of activities. Adapting their structure may require changes to the underlying activities. As many activities are reused in multiple processes, changes in one process can indirectly influence other parts of an organization and responsibilities can be blurry [70]. We consider processes with a small number of shared activity contexts ($SAC$) more feasible for optimization than processes with many $SAC$. Equation 7 presents the underlying formalism, whereby $NC(A_{ijk})$ denotes the number of unique contexts in which an activity occurs.

$$SAC_i = \left( \frac{\sum_{j=1}^{|P_i|} \sum_{k=1}^{|C_{ij}|} NC(A_{ijk})}{\sum_{j=1}^{|P_i|} |C_{ij}|} \right)^{-1} \quad with \quad NC(A_{ijk}) = \left| \bigcup_{\hat{\imath}}^{|E|} \bigcup_{\hat{\jmath}}^{|P_{\hat{\imath}}|} \bigcup_{\hat{k}}^{|C_{\hat{\imath}\hat{\jmath}}|} \left\{ A_{\hat{\imath}\hat{\jmath}(\hat{k}-1)}, A_{\hat{\imath}\hat{\jmath}\hat{k}}, A_{\hat{\imath}\hat{\jmath}\hat{k}} \right) : A_{\hat{\imath}\hat{\jmath}\hat{k}} = A_{ijk} \right\} \right|$$

Equation 7: Shared Activity Contexts

We further assess the adaptability of a process based on the number of stakeholders involved in its execution. In general, a higher number of involved stakeholders influences the adaptability of a process negatively [11, 71]. To facilitate stakeholder adoption, companies rely on change management, which demands a communication plan, project management, and incentives. This is mostly due to a resistance to change in workflows, tasks, and responsibilities. Thus, we consider the number of involved stakeholders ($S$) as an indicator for process feasibility. Equation 8 presents the underlying formalism, whereby $Res(A_{ijk})$ denotes the resource executing activity $A_{ijk}$.

$$S_i = \left(\left|\bigcup_{j=1}^{|P_i|}\bigcup_{k=1}^{|C_{ij}|} Res(A_{ijk})\right|\right)^{-1}$$

Equation 8: Stakeholder Involvement

Lastly, the feasibility to improve a process also depends on the sheer size of the process, that is the number of activities in a process, which are characterized by interdependencies regarding work and information flows. Adapting processes with many activities requires more effort for analyzing these flows and for maintaining a workable process design. We use the average length of a process ($L$) as an indicator for process feasibility. Equation 9 presents the underlying formalism.

$$L_i = \left(\frac{\sum_{j=1}^{|P_i|}|C_{ij}|}{|P_i|}\right)^{-1}$$

Equation 9: Process Length

## 5 Analysis of Case Company Data

Having defined an indicator system to measure the importance, health, and feasibility of processes, we subsequently analyze the case data. First, we introduce the companies and describe their main characteristics. Second, we detail the employed procedure for process discovery, compute the indicators, and examine their distribution at company level to answer RQ2.

### 5.1 Descriptive Statistics on the Case Companies' Processes

The case companies operate within the sectors of financial services and public administration. As they differ regarding their size and business focus, each event log contains information about different processes and its activities. Because all logs capture a random time period, we preprocess the data prior to process discovery by cleansing events traces that were started before or completed after recording began where feasible.

#### 5.1.1 The IT Service Management Log

The first event log stems from a Dutch multinational banking and financial services company, which offers services related to banking, insurance, leasing, and real estate [48]. More specifically, the event log is recorded in the company's IT department and contains procedures linked to the management of customer inquiries by service desk employees. IT service management is a common and comparable process in any IT-enabled company. We have preprocessed the log and removed incomplete traces not containing both, an *open* and a *closed* event. The log covers the time period between January 2013 and March 2014 and contains 46,146 traces with a total of 463,487 events. While the shortest trace consists of 2 events, the longest includes 178 events. In addition to a unique identifier, each event contains attributes that specify its status, involved teams, urgency, priority, open and closing time, related interaction, and related change. We illustrate the log's main characteristics in Figure 4.

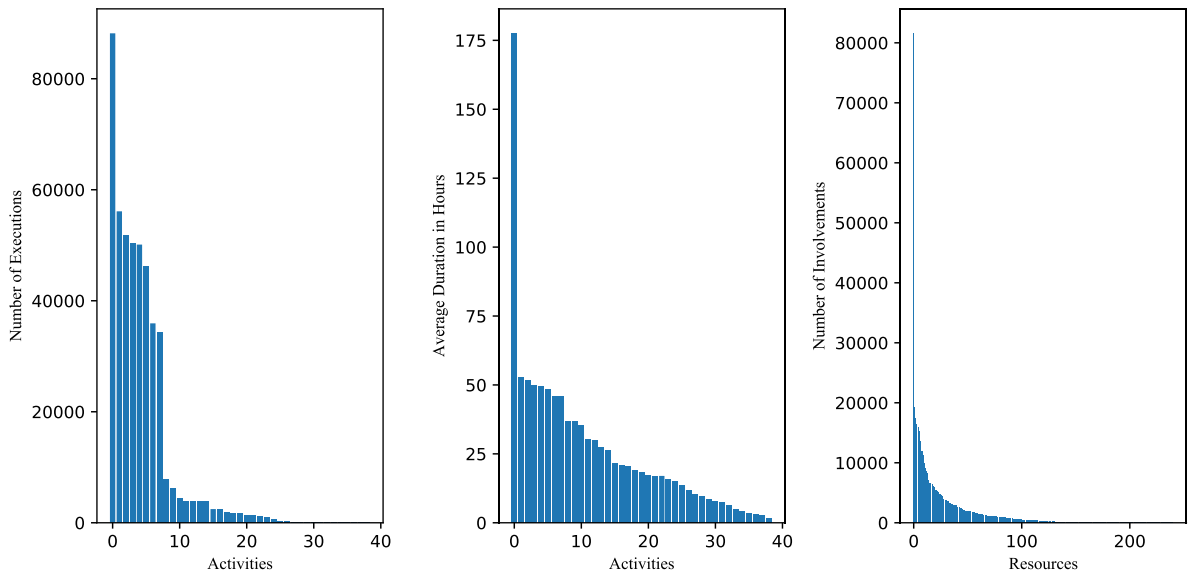

Figure 4: Main Characteristics of the IT Service Management Log

As shown in the first chart of Figure 4, the company's IT service management process consist of a few activities with a high execution frequency, while most activities are rarely performed. With over 86,000 executions, the activity of *assignment* constitutes the most frequent activity, followed by *operator update* with 55,000 executions. The second chart illustrates the average duration of each activity. The shortest task takes approximately one hour. While most activities take less than 25 hours, the most complex tasks take up to 50 hours of working time with an outlier at 175 hours. Ultimately, the third bar chart summarizes the frequency of team involvements in process executions. *Team 0008* performs over 80,000 activities and represents the busiest resource. By contrast, 27 teams participate in only one or two process executions.

### 5.1.2 Loan Application Log

We collected two event logs from at a Dutch financial institute, which describes applications for personal loans or overdraft [49, 50]. The first log covers the comparably short time period between October 2011 and March 2012 and contains 13,087 cases with roughly 262,200 events. However, the second event log comprises all events occurring in 2016 containing 31,509 cases and roughly 1,200,000 events. It was recorded with a new workflow system and advice from the analysis of the 2012 log has been incorporated. In addition to activities necessary to handle the application themselves, the logs also provide information about offers sent out to customers and work items that belong to the process. We summarize the log's main characteristics in Figure 5. All figures relating to the loan application log feature two curves for the 2012 (blue) and the 2017 (orange) data.

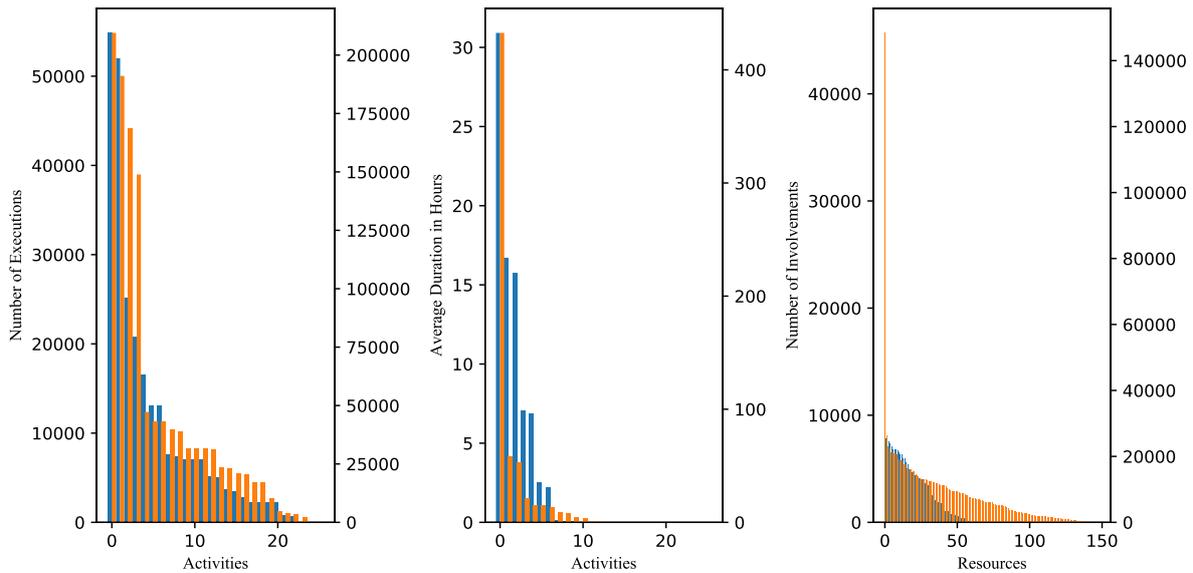

Figure 5: Main Characteristics of the Loan Application Log

The first chart of Figure 5 illustrates the distribution of activity executions. While the 2012 log contains 24 unique tasks, *completing customer request* and *list offers* are executed more than 50,000 times and represent the most frequently performed activities. With 12 executions, *modify contract details* is the least frequently performed activity. The second chart summarizes the average duration of activities within the covered time period. With an average completion time of 30 hours, the task of *validate request* is the most time-consuming activity, followed by *modify contract details* with 37 hours. The duration of most tasks is neglectable, as their completion time lies below ten seconds, which is primarily due to their automated and rule-based execution. Ultimately, we present the distribution of involved resources in the third chart. All traces in the event log start with the activity *submitted*, which is always executed by resource *112*. As a consequence, it is involved in approximately 45,000 activity executions and thus the busiest organizational resource. Most other resources execute between 1,000 and 7,000 activities, while a small group of 9 resources performs less than 100 tasks.

The 2017 log contains 26 unique tasks, *validate application*, *call after offers*, *call incomplete files*, and *complete application* are the most frequent tasks with more 148,000 to 209,496 executions. With 22 executions, *personal loan collection* is the least frequent activity. The task of *personal loan collection* is the most time-consuming activity with an average activity duration of over 432 hours. The duration of many tasks is neglectable as well. All traces in the event log start with the activity *create application*, which is executed by resource *User_1* in two thirds of the cases. As a consequence, it is involved in approximately 150,000 activity executions and thus the busiest organizational resource. Most other resources execute up to 20,000 activities, while a small group of 8 resources performs less than 100 tasks.

### 5.1.3 Building Permit Application Log

The third event log stems from five Dutch municipalities concerned with building permit applications [51]. Building permits are required for all projects with an impact on land or environment ranging from new construction projects over building demolishment to cutting a tree. The recorded cases contain information about the main application and objection procedures in various stages. The log was recorded from June 2011 to January 2014 and consists of over 5,600 cases and 262,600 events. In general, the log contains a comparably large number of activities and resources and is rather complex due to many different configurations and variants. We illustrate the log's main characteristics in Figure 6.

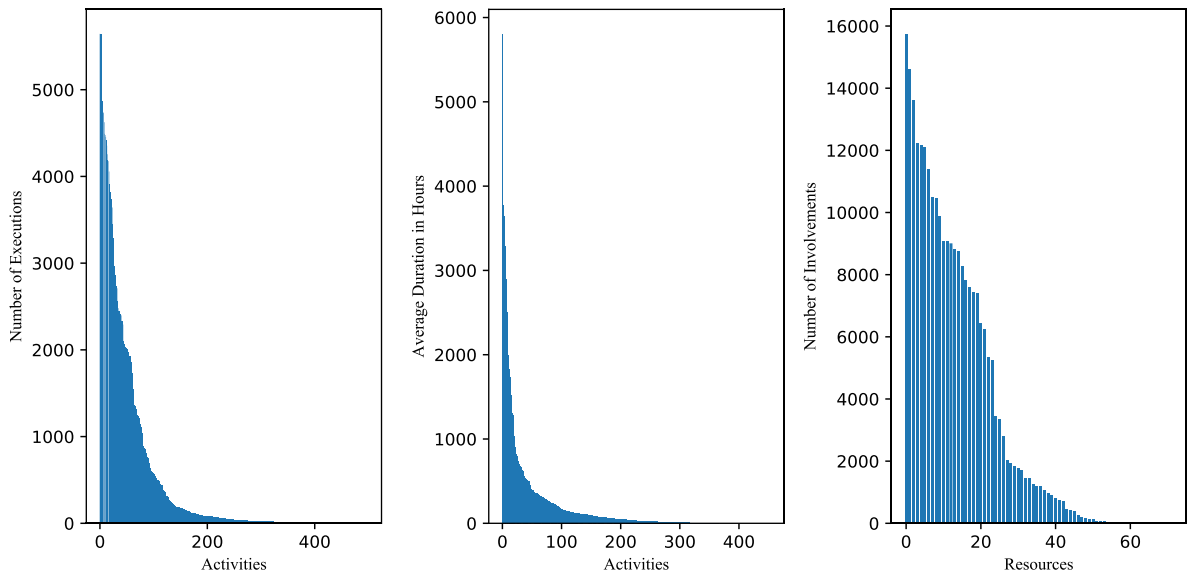

Figure 6: Main Characteristics of the Building Permit Application Log

The first bar chart in Figure 6 illustrates the distribution of execution frequencies for the activities contained by the event log. The log contains a total of 460 unique tasks. Corresponding frequencies range between 1 and 5,600, while most activities are performed up to 1,000 times. Regarding their average duration, the most time-consuming activities take up to 5,800 hours (240 days), while the quickest tasks are executed automatically and require no significant waiting or working time. Most resources are frequently involved in the company's process variants. More specifically, resource *560781* accounts for 15,748 executions and represents the busiest resource. By contrast, resource *560427* is only involved in two activity instances.

### 5.2 Process Discovery

In order to examine the distributions of the performance indicators, it is necessary to merge similar cases into process variants for handling business objects. We use agglomerative clustering as outlined in Section 3.2. The introduced selection criteria yielded ambiguous or contradictory results that range from two clusters to far more than 1,000 clusters and hardly supported decision making. This is mostly due to the sparsity and dimensionality of the distance matrices and due to a significantly uneven distribution of feature counts [72]. Hence, we draw upon the resulting dendrograms and the within-cluster sums of squares to select an appropriate number of clusters. We illustrate the dendrograms in Figure 7.

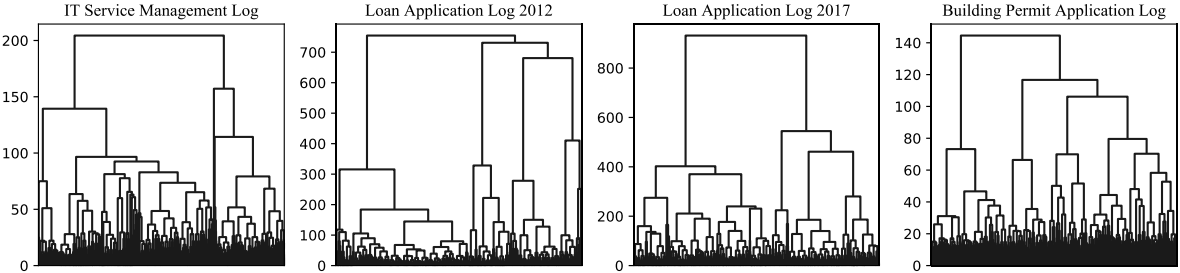

Figure 7: Cluster Selection Based on Dendrograms

In the case of the IT service management log, the dendrogram starts out with roughly 45,000 clusters. Judging from the relative distance increase, we selected a 150-cluster solution for further analysis. As shown in Figure 7, the loan application log has less unique traces but is characterized by an increase of dissimilarity when merging clusters at low levels. Consequently, we decided on a 200-cluster solution. Ultimately, the building permit application log contains more unique traces and, thus, starts out with more clusters at the bottom of the dendrogram. However, merging clusters does not result in large increases in dissimilarity, justifying a selection of 300 clusters. We have calculated the distributions in Section 5.3 with multiple cluster sizes to evaluate the effect of different cluster sizes on the analysis. As a result, we only noticed a smoothing of the curves rather than other emerging distributions. As expected, a long tail is not as discernible for very small cluster sizes below 50.

## 5.3 Distributions of the Process Performance Indicators

With the event log broken down into homogenous process variants, we can measure and visualize the indicators of our indicator system. We apply the predefined formalisms to each process and aggregate them at company level. For visualization purposes, we have normalized the scales for all indicators on the *y*-axis from 0 to 1. The *x*-axis shows the process variant clusters ordered by their *y* value.

We visualize the distributions of the nine indicators for the IT service management log in Figure 8.

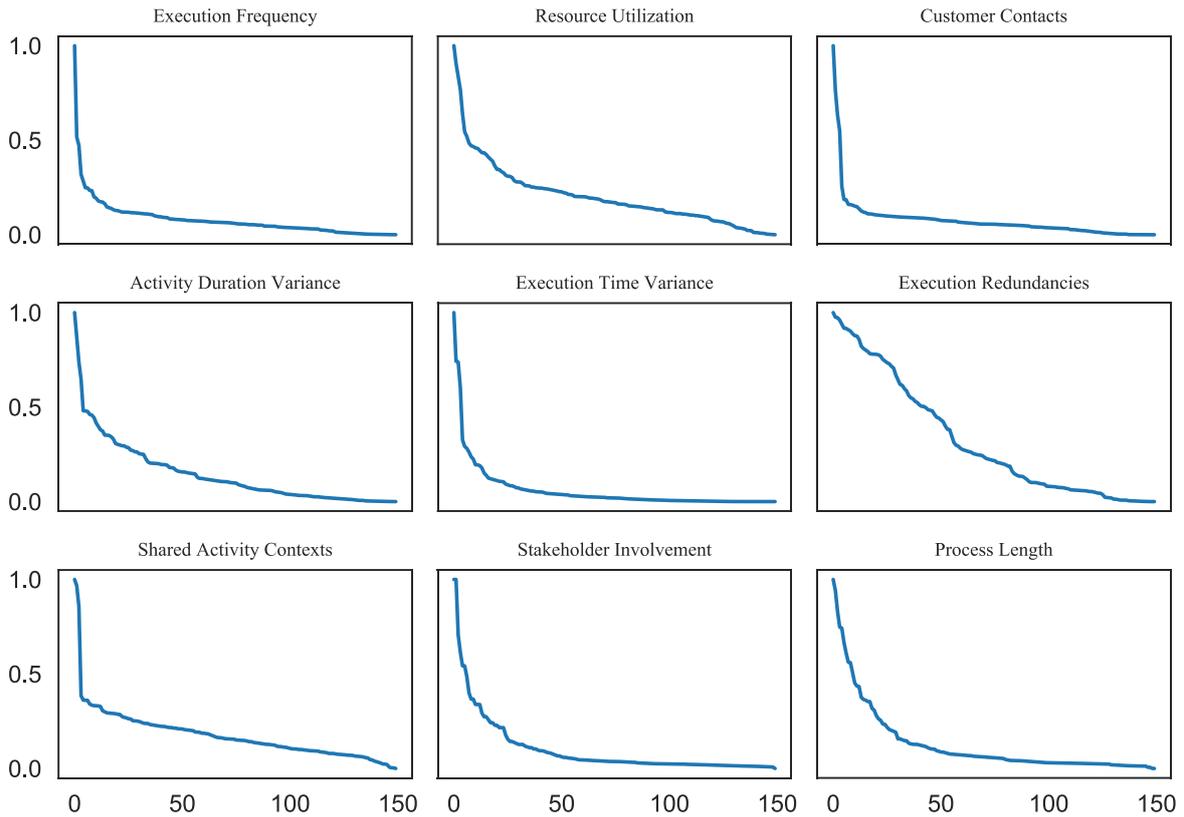

Figure 8: Indicator Distribution for the IT Service Management Log

As shown in Figure 8, most indicators are characterized by an exponential distribution. In particular, only a few process variants have a high number of *customer contacts*, while others involve little to no customer interactions. Hence, only a handful of process variants influence customer satisfaction directly and are, thus, considered as important in that respect. Furthermore, most process variants show little deviations in *execution time variance*, which indicates a high degree of standardization and structuration. Only ten process variants exhibit significant deviations. We consider these ten process variants as particularly interesting concerning the process's health. They should be scrutinized and possibly harmonized or reduced as on the one hand, they may be dysfunctional. On the other hand, these deviations could be natural from variability in the demand. That is, some more complex IT service management cases naturally take longer than simpler requests. Nevertheless, at the very least the variant process models should be revisited to understand where this variance comes from. Lastly, the indicator of *shared activity contexts* and *resource utilization* imply that most process variants are not overly connected or include important stakeholders. It is, however, less prominent in comparison to the other indicators. Summarizing, we can summarize that the values of most indicators are distributed in an exponential manner. Only *execution redundancies* clearly diverge from this observation. Only a few process variants contain only a few loops, while most of them are characterized by many and an almost linearly decreas-ing number of loops. It indicates a less clear-cut differentiation between high-value and low-value process variants. It is not entirely surprising as IT service ticketing processes often solve a problem iteratively.

Figure 9 summarizes the indicator distributions for the loan application log. The blue line represents the 2012 log, the orange line visualizes the 2017 log.

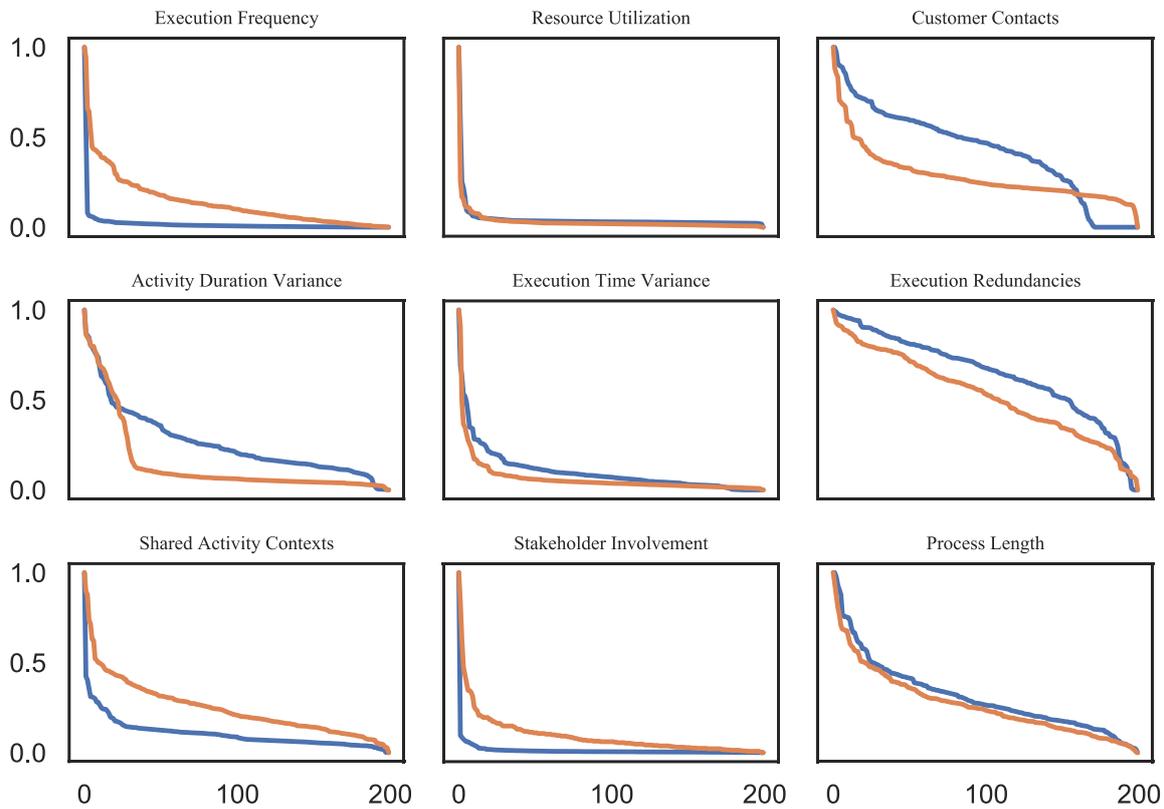

Figure 9: Indicator Distribution for the Loan Application Log

As shown in Figure 9, the indicators of *execution frequency*, *resource utilization*, *execution time variance*, and *stakeholder involvement* have a similarly exponential distribution in particular for the 2012 dataset. This implies that regarding the respective indicators very few process variants are extraordinarily important, unhealthy, or feasible to change. By contrast, most process variants are characterized by a significantly lower improvement potential. The indicators of *stakeholder involvement, resource utilization,* and *execution frequency* show an exceptionally steep function profile. This suggests that only a very limited amount of process variants requires the involvement of many teams or are executed on a regular basis. The 2017 dataset appears to be more balanced in this respect with *execution frequency* and *stakeholder involvement* being less steep. The assumption of a strict long tail distribution does not hold for the indicators of *customer contacts* and *execution redundancies*. The distribution of *customer contacts* begins with a strongly decreasing part (especially for 2017), followed by an area that runs toward the x-axis. Finally, small number of process variants involves very few teams, resulting in another steep drop of the distribution. Regarding *execution redundancies*, our results suggest that a many process variants have redundant activities that decrease linearly across the company's organizational structure. Unlike other indicators, we can also observe few process variants with few redundancies indicating a possibly inefficient process. Interestingly, the number of execution redundancies has increased slightly from 2012 to 2017. This entails that removing (unnecessary) loops was not part of the optimization or to the contrary that more checks have been implemented. On the positive side, activity duration variance has decreased and reuse in terms of shared activity contexts has increased from 2012 to 2017. In general, the results of the two datasets are comparable. We assume the 2017 curves to be smoother and less steep also due to the completeness of the dataset, which provides a more comprehensive picture of the process.

Finally, Figure 10 summarizes the indicators for the building permit application log.

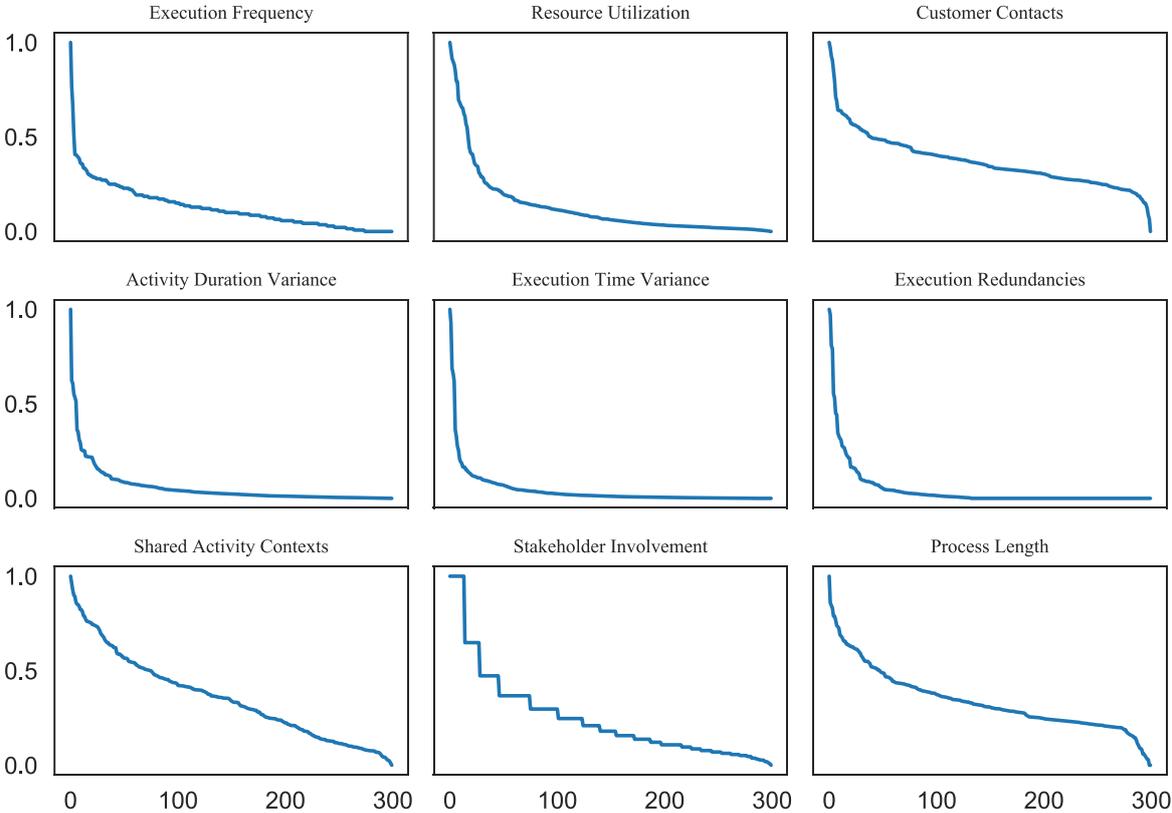

Figure 10: Indicator Distribution for the Building Permit Application Log

As shown in Figure 10, most indicators applied to the building permit application log also exhibit exponential value distributions. It applies to the indicators of *execution frequency, resource utilization*, *activity duration variance*, *execution time variance*, *execution redundancies*, and less prominently to *customer contacts* and *process length*, which both feature a sharp drop at the end of the curve indicating a very small fraction of incredibly short variants or with minimal customer contacts while most processes have a fair amount of both except for the short head. Notably, the indicator of *stakeholder involvement* forms multiple plateaus, indicating several process variants involving an equal number of stakeholders, which may hint at silos. The assumption of a long-tailed distribution does not hold for the indicator of *shared activity contexts*, which may imply many interdependencies in the process's structure. Ultimately, we consolidate using the normalized squared averages across all nine indicators per process variant (cluster) to obtain an integrated value for the overall prioritization of variants for improvement of each the companies' processes. Figure 11 illustrates the corresponding distributions for the four event logs of the three case companies sorted by the most valuable variant.

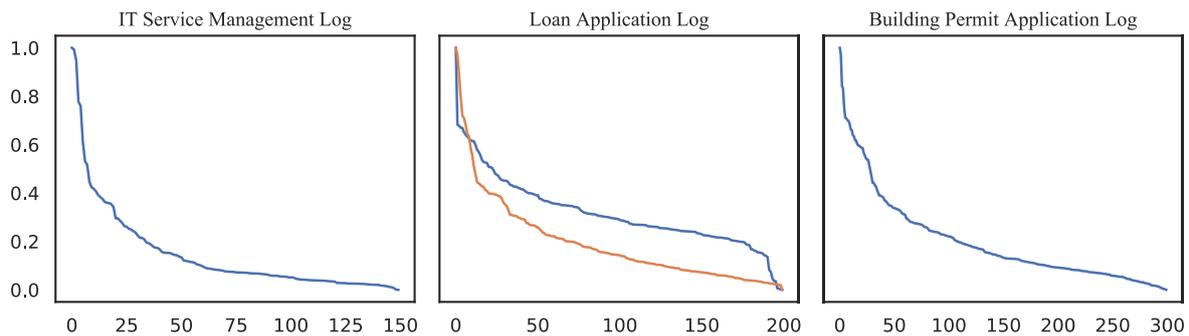

Figure 11: Indicator Distributions at Process Level

All case companies show an exponential distribution of process improvement potential that can be used for process variant prioritization. Hence, a few process variants are characterized by high values across the indicators of our indicator system. These process variants form the short head of the companies' process variant distributions. Beyond that, a long tail contains process variants that are either less important, healthy, or feasible for adaptation. While not all individual indicators resulted in long tail distributions, their integrated value clearly shows that process value and, thus, improvement potential is distributed in an exponential long tail curve. The loan application log curve for 2012 is shift up with a less prominent long tail indicating not only a short head and a long tail but also a short tip of the tail with process variants with comparably low score in contrast to the average variant. These were probably revisited before the new system was implemented that is used in the 2017 log, which shows a long tail curve.

In sum, this substantiates Imgrund, et al. [11]'s assumption of the long tail of business processes on the microscopic level of a single processes. The generalization of the analysis on the macroscopic level of all processes in a company is conceivable and we have defined the indicators in a way that they could be used for this task if the data was available to us.

## 6  Discussion and Implications

In this section, we discuss our results and their implications for BPM. Specifically, we determine and examine the composition of the microscopic long tail of business processes of our case companies' processes and analyze the characteristics of the process variants located in the short head and the long tail of the process distributions to offer further insights and recommendations. Thereby, we answer RQ3.

## 6.1 Analyzing the Short Head and the Long Tail

In this study, we analyzed process execution data from three case companies. This entailed merging event log cases into process variants, assessing a total of nine improvement indicators, and aggregating them at process level. We found that the aggregated distribution of improvement potential is exponential (cf. Figure 11) and results in a situation where companies face a long-tailed curve with a short head and a long tail of business processes or business process variants. These two areas are separated by a *line of manageability,* whose location depends on situational circumstances, for example available resources or skills [11]. In our research, we separate long tail processes from those located in the short head based on the Pareto principle (i.e. the top 20 % process variants for the short head), which is the fundamental assumption of long tail economics [10]. Companies with a lesser BPM skillset or limited resources may only consider only a handful of processes in the short head instead.

In order to assess the composition of the short head and long tail, we have calculated how important each indicator was for classifying process variants as either short head or long tail. That is, we calculated the percentage from the aggregated total value of each indicator for the process variants residing in the short head and process variants residing in the long tail. As introduced above higher values in the dimension of feasibility (i.e. *shared activity contexts*, *stakeholder involvement*, and *process length*) suggest lower absolute counts and vice versa. Figure 12 provides insights into the individual distributions.

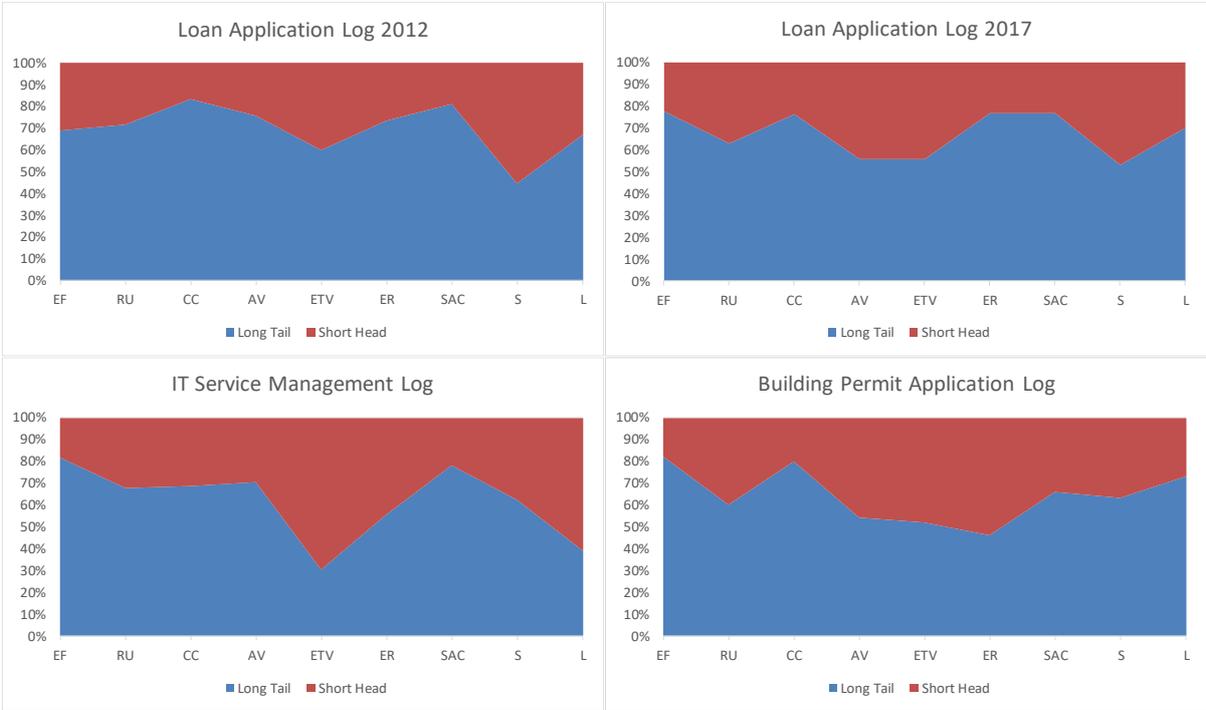

Figure 12: Contribution of Process Characteristics

For the IT service management log, *execution time variance* and *process length* were clearly stronger indicators for the short head, while high *execution frequency* and low *shared execution contexts* point towards the long tail. This entails that there were some comparably short process variants that varied widely in their execution time, while the duration of individual activities did not vary greatly. Further, there variants were not executed often and did not share activities much.

In the loan application logs, *stakeholder involvement* and *execution time variance* are characteristics of the short head of processes. For the 2017 log, *activity duration variance* and *resource utilization* are noticeable as well, which were weak indicators for the long tail in 2012. The key aspects of the long tail with about 80 % for the relative value in both logs are *customer contacts* and *shared execution contexts*. In the 2017 log, *execution redundancies* and *execution frequency* increasingly point to the long tail as well.

Lastly, in building application permit log we separated the short head and the long tail as follows: *execution redundancies* and to a lesser extent *execution time variance* and *activity duration variance* point to the short head, *execution frequency*, *customer contacts*, and *process length* point to the long tail.

Figure 13 visualizes the aggregated contribution across all four logs as an averaged total. This allows us to assess which indicators had the most impact across all companies.

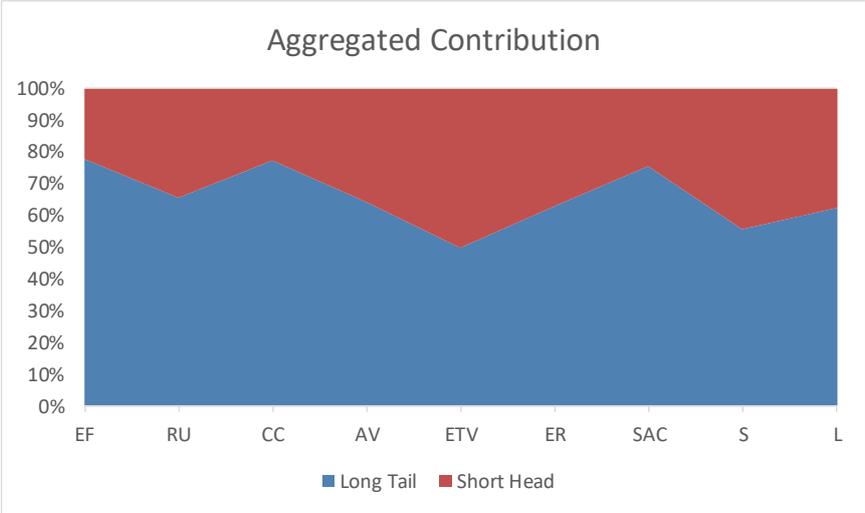

Figure 13: Aggregated Contribution of Process Characteristics

Relatively speaking, short head process variants of our case companies showed a low *stakeholder involvement* as well as a high *execution time variance*. Process variants in the long tail showed high *execution frequency*, *customer contacts*, and low *shared activity contexts*. Prioritizing process variants that have relatively low *stakeholder involvement* to reduce their *execution time variance* is intuitively comprehensible and positions process variants in the short head that are have a bad health and are feasible to improve. Few *shared activity contexts* mean less complexity as they do not touch many processes. Interestingly, a high number of *customer contacts* does not point to the short head but the long tail. This may be due to the nature of the selected logs which are limited in their customer-facing nature. While a higher *execution frequency* describes process variants as important due to their high volume of transactions, it is apparently not as important as *resource utilization* to identify the short head of processes. While this is surprising at first glance, in hindsight it supports the current trend of automating processes of the middle part of the long tail curve using robotic process automation [12].

As no indicator clearly dominated the indicator system, we refrain from formulating propositions on distinct universal characteristics of the short head or long tail of processes other than stating that due to the balanced nature of our indicator system, the composition will be most influenced by situational conditions in most cases. We acknowledge that this may also be due to our process logs, which do not originate from large multi-national corporations. Processes run in 50 countries across languages and geographies may convey a clearer picture.

However, in conformance with long tail theory, our long tail accounts for 80 % of all process variants and contributes about two thirds of the indicator values rather than the assumed 20 % following the Pareto principle [10]. This underscores their practical importance, not only financially but also for example in terms of personal content with the process organization due to functional rather than dysfunctional processes.

In summary, our data shows that deriving a general picture of the short head and long tail of processes – at least on the microscopic level – is difficult and may be even specific to the application area or industry. In addition, as Anderson [10] already noted, moving the line of manageability up and down the curve may result in somewhat different results. Our observations should be interpreted in this light.

**6.2  Implications for a Hybrid Approach to BPM**

Nowadays, many BPM initiatives are organized centrally. This entails identifying and implementing improvement potentials in a top-down manner [7] for the short head of business processes only. Typically conducted by a few process analysts, central approaches yield partial improvements due to resource constraints and the inherent need for process prioritization [7]. By contrast, a hybrid approach would imply also using decentral initiatives employing IT for communication, coordination, and collaboration to connect distributed stakeholders, who realize incremental and continuous improvements jointly to improve a company's overall performance [11]. The applicability of hybrid BPM depends on company and process characteristics as well as prioritization criteria.

In the following, we map them to the properties of the case companies' short head and long tail across the analyzed processes based on our indicators and present some observations. These primarily apply to the process variants, we discussed above but some observations seem generalizable and in a larger study could be confirmed or expanded to a macroscopic level.

According to our data logs, we found process variants that are feasible to optimize in the short head, which allows for efficient and less complex central improvements projects for example due to a low number of involved stakeholders. Dysfunctionality in terms of varying process cycle times may indicate demand variability, complexity, or frequently changing requirements, leading to situations in which these process variants lie idle or must be executed multiple times. Hence, optimizing short head process variants is likely to require expert knowledge as well as advanced tools and techniques. These process variants demand a careful and periodic redesign by specialized process analysts. Thus, companies can account for their direct impact on customer satisfaction and reduce the risks of unforeseen events and a loss of reputation.

In our case, process variants in the long tail comprise more stakeholders, which makes them difficult to optimize in decentral initiates. Hence, it is important to focus on incremental improvements. As argued above, with robotic process automation there is a means to automate frequently executed activities in these process variants in a lightweight manner even for decentral initiatives. Less shared activities make this a less complex endeavor than for many long tail processes. The same applies to those activities with customer contacts. Nevertheless, an incremental approach enables distributed stakeholders to identify and implement process adaptations gradually, without an elevated risk of reducing customer satisfaction or hampering business performance. Their participation and a constantly increasing process awareness further provide a suitable basis for the collection of relevant process and customer information. Current frameworks do not yet reflect this [cf. e.g. 73].

In line with long tail theory, despite their smaller individual optimization potential, in sum, processes in the long tail yield substantial joint continuous improvement potential in the long term. In purely centralized BPM initiatives, these processes or process variants would most likely not be identified as suitable optimization candidates, as expected benefits fall short of corresponding costs. In support of decentral initiatives, companies should emphasize the importance of suitable training offers and continuous monitoring and help by the central initiative.

## 7    Conclusion, Limitations, and Future Research Opportunities

In this paper, we have scrutinized the theory of the long tail of business processes by analyzing process execution data from three case companies with different characteristics and business foci. Due to the lack of generally applicable measures for process prioritization based on event log data, we have defined an indicator system and thereby answered RQ1.

The indicator system draws upon the high-level criteria of importance, health, and feasibility and are reusable for most available event log configurations. We employed the indicators to each event log and discovered that their values generally follow a long tail distribution for the variants of a single process. Furthermore, we found that the long-tailed shape also holds as an aggregate of all indicators at process level. This implies that the case companies perform a few high-value processes variants with large improvement potential that together form the short head of their process distribution. Beyond that, a long tail contains process variants, whose improvement potential is neglectable if managed centrally. Thereby, we answered RQ2 and validated the general assumption of the long tail of business processes at the microscopic level. We also provided some thoughts on the transfer to the macroscopic long tail of processes.

To answer RQ3, we examined constitutive characteristics of process variants located within both regions of the companies' process variant distribution and found some emerging patterns. Nevertheless, we have noticed enough variance across the indicators to conclude that no indictor dominates the short or the long tail of processes, which on the one hand confirms the indicator system to be balanced, but on the other hand impedes drawing clear boundaries or establishing a line of manageability.

Due to the fresh perspective of our research, we provide novel insights related to the findings of prior BPI challenges. While we do not provide significant insights into topics related to our log preprocessing, namely process clustering, we are the first contribution to provide a balanced set of indicators across the dimensions importance, health, and feasibility that can be used to prioritize process and their variants based on event log data. In this way short head process (variants) can be analyzed systemically and improved by a central BPM initiative that builds upon expert knowledge and a strong connection to a company's overall objectives and strategies. Long tail process (variants), on the other hand, should be distributed for incremental improvement efforts, which are organized and implemented by stakeholders at their place of execution. We are the first, to provide insights into the composition of the highest value cases versus the long tail of cases. Under the assumption of ever-increasing log sizes, our contribution may support the preprocessing of event logs to focus on the most relevant variants for detailed analysis.

However, this research is not without limitations. First, due to a limited data availability, we have only examined four process logs from three Dutch case companies. The event log data does not cover the companies' entire organizational structure and process landscape but contains traces and events from a single process each. While we are convinced that our microscopic results provide generalizable implications for the long tail of processes and apply to other business contexts as well, more and larger studies with company level event logs are necessary to provide more definitive evidence. Second, our indicator system is not without limitations. Although we analyzed the literature for performance indicators that are suitable for prioritizing processes based on execution data, we mostly relied on combining existing knowledge and deducing concepts from empirical observations. Consequently, we cannot completely eliminate the risk of having neglected indicators that might have offered additional insights or research outcomes. Third, this research strongly builds on exploratory methods, which frequently lack validity and reliability. We provide a detailed description of the conducted research procedure to tackle this shortcoming. However, different configurations of the trace clustering approach might yield different cluster configuration. We have tried to alleviate this issue by reviewing the curves of multiple cluster sizes for each company. Similarly, and fourth, moving the line of manageability in any company-specific case may result in different compositions of the short head, which may be more related and relevant to the circumstances of the company. Addressing the drawback requires mixed-method studies, which allow for a continuous evaluation of computational results by collecting and consolidation real-world insights. This represents a future research opportunity.

Further opportunities lie in the evaluation of the proposed indicator system and in applying it to a larger number of event logs from one or more companies. Further insights may be gained by incorporating weighing factors for the indicators to prioritize within the indicator system according to the company's goals. Moreover, it is necessary to obtain a more detailed understanding of the characteristics of processes located in the long tail and short head of the process distributions of companies. Resulting implications can then be consolidated and used as a foundation for the definition of a comprehensive management framework to collect and analyze the data but also to provide a qualitative perspective, which should always be considered before making decisions.